\documentclass[aps,prc,twocolumn,times,graphicx,tighten,showpacs]{revtex4}
\usepackage[dvips]{graphicx}

\newcommand{\bea}{\begin{eqnarray}}
\newcommand{\eea}{\end{eqnarray}}
\newcommand{\be}{\begin{equation}}
\newcommand{\ee}{\end{equation}}

\begin{document}

\title{
Nuclear model effects in Charged Current neutrino--nucleus 
quasielastic scattering}

\author{C. Maieron, M.C. Mart\'{\i}nez and J.A. Caballero}

\affiliation{ 
Departamento de F\'\i sica At\'omica, Molecular y Nuclear \\ 
Universidad de Sevilla, Apdo.1065, E-41080 Sevilla, SPAIN}

\author{J.M. Ud\'{\i}as}
\affiliation{
Departamento de F\'\i sica At\'omica, Molecular y Nuclear \\ 
Universidad Complutense de Madrid, E-28040 Madrid, SPAIN}

\date{\today} 


\begin{abstract}
  The quasielastic scattering of muon neutrinos on oxygen 16
  is studied for neutrino energies between 200 MeV
  and 1 GeV using a relativistic shell
  model. Final state interactions are included within the
  distorted wave impulse approximation, by means of a relativistic
  optical potential, with and without imaginary part, and of a relativistic
  mean field potential. For comparison with experimental data
  the inclusive charged--current quasielastic
  cross section for $\nu_\mu$--$^{12}C$ scattering in the kinematical
  conditions  of the LSND experiment at Los Alamos is also presented and 
  briefly discussed.
\end{abstract}


\maketitle

In the past few years the observation of neutrino oscillations at
Super--Kamiokande~\cite{SK} and the subsequent proposal and realization
of new experiments, aimed at determining neutrino properties with high
accuracy~\cite{nuexp}, have renovated the interest towards neutrino
scattering on complex nuclei. In fact, neutrino detectors usually contain
carbon or oxygen nuclei, and for a proper interpretation of
the experimental results the description of the $\nu$--nucleus interaction
must be accurate~\cite{Nuint01}.
 
At intermediate neutrino energies, ranging from some hundreds MeV to a few
GeV, $\nu$--nucleus quasielastic scattering has been studied within
several approaches~\cite{medium}. Relativistic and non relativistic
studies of Random Phase Approximation have shown nuclear
structure effects to be relevant only at low momentum transfers, but
indications have been found that for future and precise data analyses of,
e.g., atmospheric neutrino measurements, more accurate theoretical
estimates may be needed. Additionally, very recently new attention has been
drawn towards final state interaction (FSI) effects, which, contrary to what is
often assumed, may still be relevant even at the relatively high energy
$E_\nu=1$ GeV~\cite{Bleve}.

In this contribution we study charged--current (CC) neutrino--nucleus
quasielastic scattering within the framework of a Relativistic Shell
Model (RSM), already successfully employed to study exclusive electron
scattering~\cite{eeprimeN} and Neutral Current
neutrino scattering~\cite{Alberico:1997vh}. We compute inclusive
$\nu_\mu$--$^{16}O$ quasielastic cross sections for three values of the
incident neutrino energy, namely $200$ MeV, $500$ MeV, and $1$ GeV, which
are representative of the kinematical range where quasielastic scattering
gives the main contribution to the inclusive $\nu$--nucleus process.

We describe the CC quasielastic scattering of neutrinos on a
nuclear target within the Impulse Approximation (IA), assuming that the
incident neutrino exchanges one vector boson with only one nucleon, 
which is then emitted, while the remaining (A-1) nucleons in
the target are spectators. The nuclear current is assumed to be the sum of
single nucleon currents,
for which we employ the usual free nucleon expression
(see~\cite{Alberico:1997vh}) 
with the axial form factor
parameterized as a dipole with cut--off mass 
$M_A=1.026$ GeV~\cite{Bernard:2001rs},
and the states of the target and residual nuclei
to be adequately described by an independent particle model wave function.  

To describe the bound nucleon states we use Relativistic Shell Model wave
functions, obtained as the self--consistent (Hartree) solutions of a Dirac
equation, derived, within a Relativistic Mean Field approach, from a
Lagrangian containing $\sigma$, $\omega$ and $\rho$ mesons~\cite{boundwf}.  
As the single-particle binding energies determine the threshold of the
cross section for every shell, in the numerical calculations, we have used
the experimental values corresponding to the binding energies of the
different shells.

\begin{figure}[t]
\begin{center}
{\par\centering \resizebox*{.48\textwidth}{0.25\textheight}{\rotatebox{270}
{\includegraphics{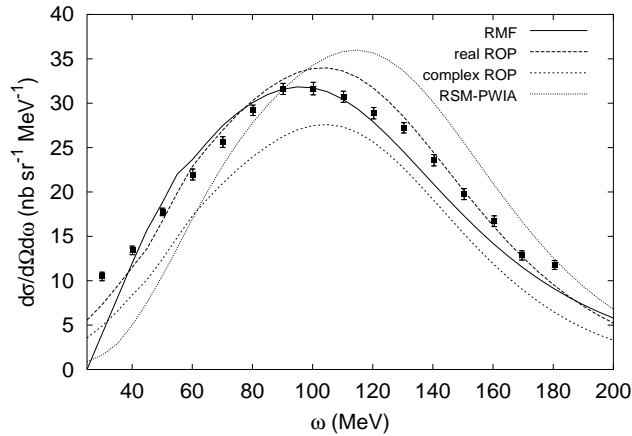}}} \par}
\end{center}
\caption{
  Differential cross section $(d\sigma/d\Omega d\omega)$
  versus the energy transfer for 
  inclusive quasielastic electron
  scattering on $^{12}C$ for a momentum transfer
  $q\simeq 400 $ MeV/c.
  The solid
  curve corresponds to the RSM with FSI described
  by the RMF, while 
  the long-dashed curve corresponds to the real 
  ROP and the short-dashed curve to the complex ROP.
  Finally, the dotted curve does not include
  FSI. Experimental data are from ref.~\protect\cite{Barreau}.}
\label{fig:sigee}
\end{figure}

For the outgoing nucleon the simplest choice is to use plane wave spinors,
i.e., no interaction is considered between the ejected nucleon and the
residual nucleus (PWIA). For a more realistic description, FSI effects
should be taken into account. In our formalism this is done by using
distorted waves which
are given as solutions of a Dirac equation containing a
phenomenological relativistic optical potential (ROP), consisting of a 
real part, which describes the rescattering of the
ejected nucleon and of an imaginary part, that accounts for the absorption
of it into unobserved channels. In this work we use the ROP corresponding
to the EDAD-1 single-nucleon parameterization presented in
ref.~\cite{Cooper}. The
use of this phenomenological ROP leads to an excellent agreement between
theoretical calculations and data for exclusive 
$(e,e'N)$ observables~\cite{eeprimeN};
however some caution should be taken in extending the conclusions drawn
from the analysis of exclusive reactions to inclusive ones. In the latter,
unless a selection of the single nucleon knockout contribution is
experimentally feasible, all final channels are included and thus the
imaginary term in the optical potential leads to an
overestimation of FSI effects.  
This is clearly illustrated in fig.~1 where we compare our theoretical results
with the experimental cross section corresponding to inclusive quasielastic
electron scattering on $^{12}C$ for a momentum transfer equal to 400 MeV/c
(similar results are obtained for 300 and 500 MeV/c).
In our calculation besides
the complex ROP, we also consider the potential obtained by setting the
imaginary part of the ROP to zero. Additionally one may
also use distorted waves which are obtained as the solutions in the
continuum of the same Dirac equation used to describe the initial bound
nucleon. We refer to this approach, 
that should be adequate to describe FSI at moderate energy transfer,
as relativistic mean field (RMF). 
As shown in fig.~1, the complex ROP results clearly underestimate the
data, while the reverse occurs for the PWIA calculation. On the other hand,
the RMF and purely real ROP agree much better with experiment, particularly
RMF for small transfer energy and real ROP for higher $\omega$.
We believe that the RMF and real ROP results
indicate a reasonable ``band'' where FSI effects should lie.

\begin{figure}[t]
\begin{center}
\includegraphics[width=0.6\textwidth]{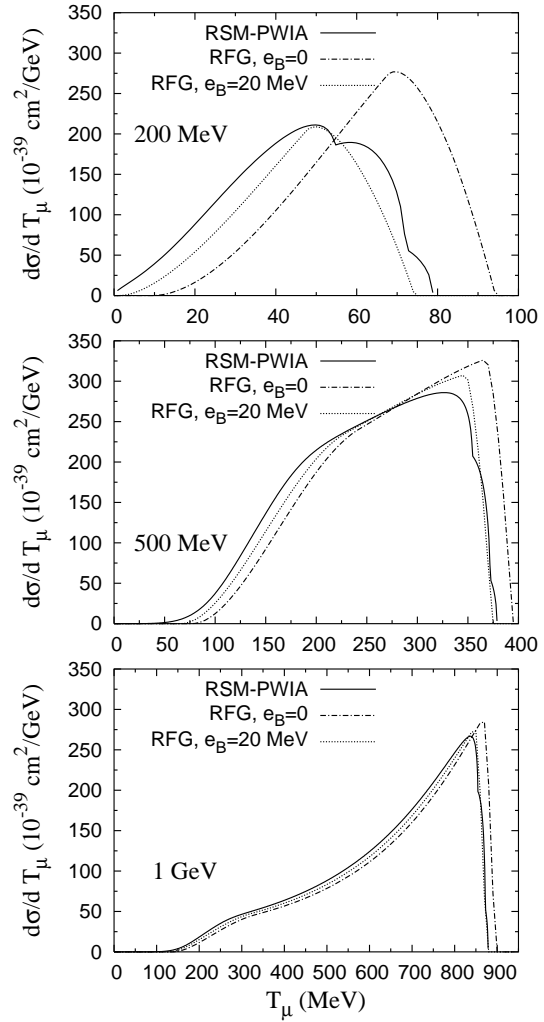}
\end{center}
\caption{
  Differential cross section $(d\sigma/dT_\mu)$
  versus the outgoing muon kinetic energy, for the quasielastic
  scattering of muon neutrinos on $^{16}O$ and for three choices of
  the incident neutrino energy: $E_\nu=200$ MeV (upper panel),
  $500$ MeV (middle) and $1$ GeV (lower panel). The solid
  curves correspond to the RSM with no final state interaction, while
  the remaining curves are calculated within the RFG, with $k_F= 225$
  MeV and $e_B=0$ (dot--dashed) and $e_B=20$ MeV (dotted).}
\label{fig:o16_dsig_pwia}
\end{figure}

Let us now present our results for neutrino scattering on $^{16}O$.
To better illustrate our model,
we start by neglecting FSI and comparing RSM--PWIA
results with inclusive cross sections obtained within the relativistic
Fermi gas (RFG)~\cite{Alberico:1997vh}. This is is done in
fig.~\ref{fig:o16_dsig_pwia}, which shows the differential cross section
$(d\sigma/dT_\mu)$ as a function of the outgoing muon kinetic energy. With
respect to the RFG curve calculated with no binding energy we observe that
the RSM cross section is reduced and shifted towards lower $T_\mu$ values,
in a way which is similar to the effect of an average binding energy in
the RFG. In addition the RSM cross section has a different shape, due to
the different momentum distributions of the single nucleon shells
contributing to the process.  Since the various shells have different
binding energies, the corresponding contributions to the cross section go
to zero at different values of $T_\mu$ and this gives rise to the
structure of $(d\sigma/d T_\mu)$ observed at large $T_\mu$.
Fig.~\ref{fig:o16_dsig_pwia} shows that
nuclear model effects on the cross sections can be rather large at low
neutrino energy, but become less relevant as $E_\nu$ increases,
practically disappearing at $E_\nu = 1$ GeV. 

\begin{figure}[t]
\begin{center} 
\includegraphics[width=0.6\textwidth]{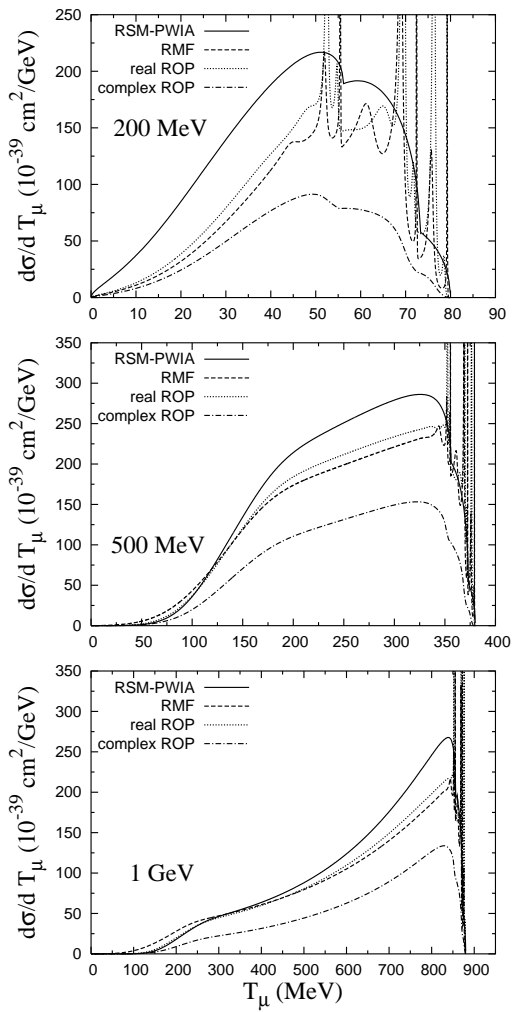}
\end{center} 
\caption{Same as fig.\ref{fig:o16_dsig_pwia}, but including
  FSI effects.  All curves are calculated within the RSM model
  in PWIA (solid), and
  within the RMF (dashed), real ROP (dotted) 
  and complex ROP (dot--dashed) approaches.
  }
\label{fig:o16_dsig_fsi}
\end{figure}

On the other hand, the behavior of FSI effects is quite different, as
illustrated in fig.~\ref{fig:o16_dsig_fsi}. Here the results obtained with
the RSM in PWIA are compared with the cases where FSI are described within
the RMF, the real ROP and the complex ROP approaches. The
use of real potentials (RMF and real ROP) for describing the final nucleon
states leads to the resonant structure observed for relatively high
$T_\mu$ (that is, small energy transfer $\omega$). 
Note that in this work we only include single-particle excitations 
within a mean field picture. Including residual interactions would make
the width and number of resonances to be considerably larger.

We observe that FSI effects
produce a reduction of the cross section, particularly important in the
case of the complex ROP model due to the absorption introduced by the
imaginary term: 
about 60\% for $E_\nu=200$ MeV and 50\% for $E_\nu=500$ MeV and $E_\nu=1$
GeV in the region close to the maximum.
For the RMF and real ROP, the reduction, similar in both cases, is
about $30\div 40\%$ for $E_\nu = 200$ MeV and $20\%$ for the other energy
values. 

\begin{figure}[t] 
\begin{center}
\includegraphics[width=0.5\textwidth]{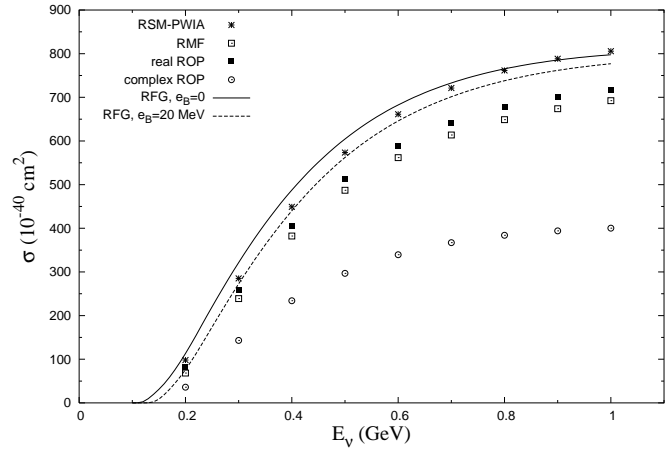} 
\caption{Integrated cross section $\sigma(E_\nu)$ for the quasielastic
scattering of muon neutrinos on $^{16}O$ as a function of the incident
neutrino energy. The curves are calculated within the RFG model with $k_F=
225$ MeV and binding energy $e_B=0$ (solid line) and $e_B= 20$ MeV
(dashed). The points correspond to RSM calculations without FSI (stars)
and with FSI effects taken into account within the RMF (empty squares),
real ROP (full squares) and complex ROP (circles) approaches.}
\label{fig:o16_tot} 
\end{center}
\end{figure}

Nuclear model effects on integrated cross sections are studied in
fig.~\ref{fig:o16_tot}, where the cross section $\sigma(E_\nu)$ is
plotted
as a function of the incident neutrino energy.
Here the contributions
coming from the RMF and ROP resonances have been included in the
calculation, in order to respect the completeness of the set of final
states predicted by the model. These contributions are important at
$E_\nu=200$ MeV, where they amount to about 10\% of the integrated cross
section, while at higher energies these effects are about 2\% (500 MeV)
and 1\% (1 GeV).
Again we see that within
the PWIA the discrepancy between different nuclear models is relatively
small and decreases with increasing neutrino energy. On the contrary FSI
effects remain sizeable even at large $E_\nu$.  As in the previous figure, the
imaginary term in the ROP leads to a too large reduction ($\sim
50\%$) of the integrated cross section.  The results for
the RMF and real ROP models, which are more reliable, show a 
smaller, but still sizeable ($\sim 15\%$ at $E_\nu=1$ GeV) reduction.

The results shown in figs.~\ref{fig:o16_dsig_fsi} and \ref{fig:o16_tot}
lead us to conclude that FSI effects should be carefully considered in
neutrino experiments which use oxygen based detectors. 
Analogous conclusions can be drawn for the case of a $^{12}C$
target.

\begin{figure}[t]
\begin{center}
{\par\centering \resizebox*{.48\textwidth}{0.26\textheight}{\rotatebox{270}
{\includegraphics{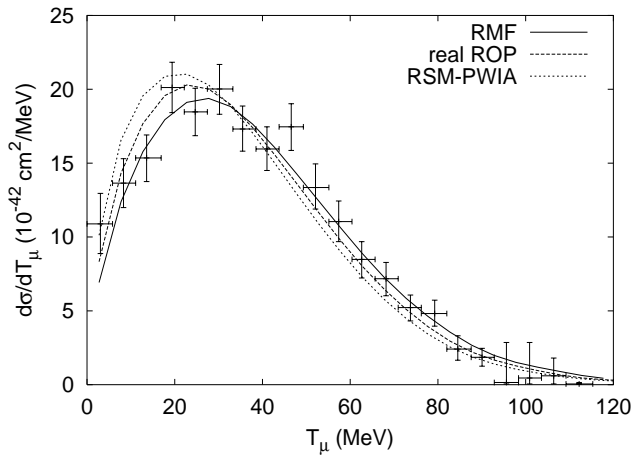}}} \par}
\end{center}
\caption{
Observed distribution of muon kinetic energies $T_\mu$
compared with the flux--averaged predictions of our RSM, in PWIA (dotted line)
and including FSI within the RMF (solid) and purely real ROP (dashed)
frameworks. 
The theoretical distributions have been normalized to give the same 
integrated values as the experimental points, 
and have been folded  in energy with a bin size of 5 MeV,
the same employed   for the experimental data.
 Data are from Albert {et al.}~\protect\cite{LSNDexp}.
}
\label{fig:dsigexp}
\end{figure}

For the purpose of comparison with the experiment,
let us consider the inclusive
$^{12}C(\nu_\mu, \mu^-)X$ cross section measured by the LSND collaboration
at Los Alamos, using a pion--Decay--in--Flight $\nu_\mu$ beam, with
energies ranging from muon threshold  to 300 MeV, and a large liquid
scintillator detector~\cite{LSNDexp}. This experiment has been
compared to very different theoretical approaches~\cite{LSNDtheory}, but an
important discrepancy still remains. Although at these low energies 
processes different from the quasielastic nucleon knockout
affect the inclusive cross section,
we consider our FSI approach to be useful to estimate
integrated cross sections.

In fig.~\ref{fig:dsigexp} we show the observed~\cite{LSNDexp} and
calculated distribution of events, averaged over the 1994 Los Alamos
neutrino spectrum $\phi(E_\nu)$ within the energy
range $E_\nu =123.1 \div 300$ MeV. The shape and position of the maximum
of the experimental distribution are approximately reproduced by the three
calculations, but the results that include FSI with the RMF 
potential are clearly
favored by the data. This is consistent with the fact that including FSI
with the mean field potential should be adequate at moderate kinetic energy 
of the ejected nucleon.
 However the values we obtain for the the flux--averaged
integrated cross section overestimate the measured cross section 
by approximately 50\%. 
More precisely, in the RSM, we obtain $\langle \sigma \rangle
= 20.5$ (PWIA), $16.8$ (RMF) and $15.1$ (real ROP) $10^{-40}$ $cm^2$.
Additional corrections due to the outgoing muon Coulomb distortion,
evaluated within the effective momentum approximation~\cite{Giusti87},
further increase these numbers by $5\div10\%$. The corresponding
final measured experimental value is 
$10.6 \pm 0.3 \pm 1.8$ $10^{-40}$ $cm^2$~\cite{LSNDexp}.

\section*{Acknowledgements}
This work was partially supported by DGI (Spain):
BFM2002-03315, FPA2002-04181-C04-04 and BFM2000-0600
and by the Junta de
Andaluc\'{\i}a. C.M. acknowledges MEC (Spain)
for a postdoctoral stay at University of Sevilla (SB2000-0427).
M.C.M. acknowledges support from a fellowship from the Fundaci\'on
C\'amara (University of Sevilla).



%

\begin{thebibliography}{99}
%
\bibitem{SK}
Y.~Fukuda {\it et al.}  [Super-Kamiokande Collaboration],
Phys.\ Rev.\ Lett.\  {\bf 81}, 1562 (1998); 
C.~K.~Jung, C.~McGrew, T.~Kajita and T.~Mann,
Ann.\ Rev.\ Nucl.\ Part.\ Sci.\  {\bf 51}, 451 (2001);
%
\bibitem{nuexp}
see for example A.~De Santo,
Int.\ J.\ Mod.\ Phys.\ A {\bf 16}, 4085 (2001);
%
\bibitem{Nuint01}
``Neutrino Nucleus Interactions In The Few Gev Region''. Proceedings, 
1st International Workshop, Nuint01, Tsukuba, Japan, December 13-16, 2001.
%
\bibitem{medium}
T.~K.~Gaisser and J.~S.~O'Connell,
Phys.\ Rev.\ D {\bf 34}, 822 (1986);
H.~c.~Kim, J.~Piekarewicz and C.~J.~Horowitz,
Phys.\ Rev.\ C {\bf 51}, 2739 (1995);
H.~c.~Kim, S.~Schramm and C.~J.~Horowitz,
Phys.\ Rev.\ C {\bf 53}, 2468 (1996), {\it ibid.}
{\bf 53}, 3131 (1996);
S.~K.~Singh and E.~Oset,
Phys.\ Rev.\ C {\bf 48}, 1246 (1993);
J.~Engel, E.~Kolbe, K.~Langanke and P.~Vogel,
Phys.\ Rev.\ D {\bf 48}, 3048 (1993);
J.~Marteau, J.~Delorme and M.~Ericson,
Nucl.\ Instrum.\ Meth.\ A {\bf 451}, 76 (2000).
%
\bibitem{Bleve}
C.~Bleve {\it et al.},
Astropart.\ Phys.\  {\bf 16}, 145 (2001);
G.~C\`o, C.~Bleve, I.~De Mitri and D.~Martello,
Nucl.\ Phys.\ Proc.\ Suppl.\  {\bf 112}, 210 (2002).

\bibitem{eeprimeN}
J.M.~Ud\'{\i}as {\it et al.},
Phys.\ Rev.\ C {\bf 48}, 2731 (1993);
{\bf 51}, 3246 (1995);
{\bf 53}, R1488 (1996);
{\bf 64}, 024614-1 (2001).
%
\bibitem{Alberico:1997vh}
W.~M.~Alberico {\it et al.},
Nucl.\ Phys.\ A {\bf 623}, 471 (1997).
%
\bibitem{Bernard:2001rs}
V.~Bernard, L.~Elouadrhiri and U.~G.~Meissner,
J.\ Phys.\ G {\bf 28}, R1 (2002).
%
\bibitem{boundwf}
C.J.~Horowitz and B.D.~Serot, 
Nucl.\ Phys.\ A {\bf 368}, 503 (1981);
Phys.\ Lett.\ B {\bf 86}, 146 (1979);
%
B.D.~Serot and J.D.~Walecka, Adv.\ Nucl.\ Phys.\ {\bf 16}, 1 (1986).
%
\bibitem{Cooper}
E.D.~Cooper, S.~Hama, B.C.~Clark and R.L.~Mercer,
Phys.\ Rev.\ C {\bf 47}, 297 (1993).
%
\bibitem{Barreau}
P. Barreau et al., Nucl.\ Phys.\ {\bf A 402}, 515 (1983).
%
\bibitem{LSNDexp}
M.~Albert {\it et al.}  [LSND Collaboration],
Phys.\ Rev.\ C {\bf 51}, 1065 (1995).
C.~Athanassopoulos {\it et al.}  [LSND Collaboration],
Phys.\ Rev.\ C {\bf 56}, 2806 (1997);
L.~B.~Auerbach {\it et al.}  [LSND Collaboration],
Phys.\ Rev.\ C {\bf 66}, 015501 (2002).
%
\bibitem{LSNDtheory}
see 
N.~Jachowicz, K.~Heyde, J.~Ryckebusch and S.~Rombouts,
Phys.\ Rev.\ C {\bf 65}, 025501 (2002);
C.~Volpe {\it et al.},
Phys.\ Rev.\ C {\bf 62}, 015501 (2000) 015501;
E. Kolbe {\it et al.}, Phys. Rev. C {\bf 52}, 3437 (1995);
E. Kolbe {\it et al.}, Nucl. Phys. A {\bf 652}, 91 (1999);
Y.~Umino, J.~M.~Ud\'{\i}as and P.~J.~Mulders,
Phys.\ Rev.\ Lett.\  {\bf 74}, 4993 (1995);
Y.~Umino and J.~M.~Ud\'{\i}as,
Phys.\ Rev.\ C {\bf 52}, 3399 (1995);
%
\bibitem{Giusti87}
C.~Giusti and F.~D.~Pacati,
Nucl. Phys. A  {\bf 473}, 717 (1987) and refs. therein.
%
\end{thebibliography}
\end{document}